\documentclass[aps,amsmath,amsfonts,amssymb,preprint,showpacs]{revtex4}
\usepackage{bm,epsfig}
\def\cC{{\cal C}}
\def\cL{{\cal L}}
\def\be{\begin{equation}}
\def\ee{\end{equation}}
\def\ba{\begin{aligned}}
\def\ea{\end{aligned}}
\def\bea{\begin{eqnarray}}
\def\eea{\end{eqnarray}}
\def\itH{\textit{Hamiltonian} }
\def\n{{\bm n}}
\def\m{{\bm m}}
\def\e{{\bm e}}
\newcommand{\had}[2]{{\hat a^\dag}_{#1,#2}}
\newcommand{\ha}[2]{{\hat a}_{#1,#2}}
\newcommand{\rn}[1]{{\hat \rho}_{#1}}
\newcommand{\bvp}[1]{{\bar \varphi}_{#1}}
\newcommand{\vp}[1]{\frac{\partial}{\partial \bar \varphi_{#1}}}
\newcommand{\bvps}[1]{{\bar \psi}_{#1}}
\newcommand{\vps}[1]{\frac{\partial}{\partial \bar \psi_{#1}}}
\newcommand{\A}[1]{\hat A_{#1}}
\newcommand{\B}[1]{\hat B_{#1}}
\newcommand{\vac}[1]{\hat v_{#1}}

\begin{document}
\title{Generating function, path integral representation, and equivalence
for stochastic exclusive particle systems}
\author{Su-Chan Park}
\affiliation{School of Physics, Korea Institute for Advanced Study, Seoul 130-722, Korea}
\author{Jeong-Man Park}
\affiliation{Department of Physics, The Catholic University of Korea,
  Puchon 420-743, Korea}
\begin{abstract}
We present the path integral representation of the generating function
for classical exclusive particle systems. By introducing hard-core
bosonic creation and annihilation operators and appropriate commutation
relations, we construct the Fock space structure. Using the state
vector, the generating function is defined and the master equation of
the system is transformed into the equation for the generating
function. Finally, the solution of the linear equation for the
generating function is derived in the form of the path
integral. Applying the formalism, the equivalence of reaction-diffusion
processes of single species and two species is described.
\end{abstract}
\pacs{05.70.Ln}
\maketitle
\section{Introduction}
Recently, nonequilibrium systems such as reaction-diffusion systems,
driven lattice systems, and cellular automata have been intensely
investigated, not only because of their connections to a variety of
important physical phenomena (nonequilibrium phase transitions,
long-range correlations, and self-organized criticality), but also
because of the analytic challenge due to the lack of a general formalism to
describe nonequilibrium systems \cite{Chopard}.
Although numerical approaches have played a major role in the
investigation of
nonequilibrium systems and have been successful in many areas, a
general field-theoretic formalism has been constantly sought because,
along with a renormalization-group analysis, it would provide a systematic
tool to evaluate physical observables in the scaling regime \cite{Marro}.
It is difficult to investigate nonequilibrium systems analytically at
the microscopic scale because there are too many microscopic variables
involved, and microscopic variables take discrete values. A
field-theoretic strategy is to find, starting from the microscopic
rules, the equations of motion of well-chosen coarse-grained
variables. These mesoscopic variables take continuous values and vary
continuously in space and time. They describe the system at the
mesoscopic scale, while keeping track of all the fluctuations related
to the microscopic degrees of freedom. 
Doi first introduced a field-theoretic formalism for
reaction-diffusion systems using the bosonic coherent state path
integral \cite{Doi}, and several authors revived the formalism and
incorporated renormalization-group approaches to the description of the
anomalous kinetics in reaction-diffusion systems and 
the stochastic sandpile model \cite{Grassberger,Peliti,Lee94,Lee95,Deem,Dickman}.

Despite the success achieved by the bosonic field theory for
reaction-diffusion systems, some nonequilibrium systems with exclusive
particles cannot be analyzed correctly by the bosonic field theory. 
Driven reaction-diffusion systems \cite{Janowsky,Ispolatov},
multispecies adsorption
models \cite{Bassler}, and driven lattice gases (DLG) \cite{Zia} 
are some examples to which the
bosonic field theory cannot be applied. In these systems, the exclusive
property of the particles is important. Several attempts have been
made to incorporate the exclusive property into a field theory. One
way to take the exclusive property into account is to exploit the
physical knowledge of the system phenomenologically as Zia and
Schmittmann \cite{Zia} have done for the DLG. This method is
an approximation and cannot be systematic. Other approaches proposed
by Brunel {\it et al.} \cite{Brunel} and Bares and Mobilia \cite{Bares}
are formulated using the
fermionic path integral and are rather difficult to analyze and to
extend for higher spatial dimensions or multispecies processes. 
Recently, van Wijland extended the bosonic field theory by introducing
the exclusion constraint operator to take the exclusive property into
account \cite{Wijland}. He used bosonic creation and annihilation operators to
construct the path integral by considering the exclusive property
using the exclusion constraint operator. 

The authors previously presented a hard-core bosonic field theory using 
hard-core bosonic creation and annihilation operators \cite{Park2000}. 
It has been successfully applied to the asymmetric exclusion process and 
several restricted solid-on-solid-type growth models to
provide the correct Langevin-type equations of motion and the proper
path integral formulation \cite{ParkGrowth}. However, the formalism assumed the
existence of the Fokker-Planck equations for processes {\it a priori} and
derived the Langevin type equations of motion.
In this paper, we develop a more general path integral
formalism for nonequilibrium systems with exclusive particles. In
what follows, we present the derivation of the path integral
formulation for systems with exclusive particles using the
generating function of the correlation functions, and we will illustrate how
the formalism can be applied to various reaction-diffusion systems to
establish nonequilibrium universality classes. 

\section{\label{Sec:formalism}formalism}
The dynamics of a stochastic system are usually described by the master
equation governing the time evolution of the probability $P(\cC;t)$.
$P(\cC;t)$ is the probability of a system being
in a microstate $\cC$ at time $t$. The master
equation takes the form \cite{G83,R84,vK97}
\be
\frac{\partial }{\partial t}P(\cC;t) = \sum_{\cC' \neq \cC} \left \{
W_{\cC \cC'} P(\cC';t) - W_{\cC' \cC} P(\cC;t)\right \},
\label{Eq:Master}
\ee
where $W_{\cC'\cC}$ is the transition rate, 
or transition probability per unit time, 
from microstate $\cC$ to $\cC'$. 
Introducing the matrix $H_{\cC \cC'}$,
\be
H_{\cC \cC'} = \delta_{\cC \cC'} \sum_{\cC''} W_{\cC'' \cC} - 
W_{\cC\cC'},
\label{Eq:Ham_elements}
\ee
Eq. (\ref{Eq:Master}) can be written succinctly as
\be
\frac{\partial}{\partial t} P (\cC;t) = - \sum_{\cC'} H_{\cC \cC'}
P(\cC';t).
\ee
Due to the resemblance of the master equation to the Schr\"odinger
equation, it is convenient to introduce Dirac's bra-ket notation and an
orthonormal basis formed by microstates $\{ | \cC \rangle \}$.
Introducing the state vector $ | P ;t \rangle \equiv \sum_{\cC}
P(\cC;t) | \cC \rangle$ and the \itH  operator $\hat H$ whose component
is $H_{\cC \cC'}$ in this orthonormal basis, the master equation can
be written in the form of an imaginary time  Schr\"odinger equation,
\be
\frac{\partial}{\partial t} | P; t \rangle = - \hat H | P ; t \rangle.
\label{Eq:master_to_Hamiltonian}
\ee
In this representation, the average of a physical quantity can be written
as
\be
\langle {\cal O} \rangle = \sum_{\cC} {\cal O}(\cC) P(\cC;t)
= \langle \cdot | \hat {\cal O} | P;t\rangle,
\ee
where $\hat {\cal O}$ is an operator with the elements
$\langle \cC' | \hat {\cal O} | \cC \rangle = \delta_{\cC,\cC'} {\cal O}(\cC)$ and
the projection state $\langle \cdot |$ is defined as
\be
\langle \cdot | \equiv \sum_{\cC} \langle \cC |.
\label{Eq:projection}
\ee

A natural way to deal with a stochastic system in which the particle
number at each site varies is to introduce a Fock-space-like
structure. When a system has $M$ different particle
species with the exclusive property, 
at each site there are $(M+1)$ possible states: a vacuum state and 
$M$ occupied states depending on particle species. 
Using Greek indices for the kinds of particle species---a 
vacuum state is regarded as a new species and 0 is reserved for 
a vacuum state---and bold Latin indices for site locations,
we denote a state at the site $\n$ by $|\alpha_\n \rangle$, where $\alpha_\n$
is the index of species occupying the site $\n$ and goes from $0$ to $M$.
For microstates of the whole system, we work in a phase space which is
composed of the direct product of single-site microstates of all sites
$\{ \n \}$,
\be
| \{ N \} \rangle = \bigotimes_{\n} | \alpha_\n \rangle.
\label{Eq:microstate}
\ee
At each site $\n$, creation and annihilation of the exclusive particles are
described by creation and annihilation operators as follows:
\be
\hat{a}_{\alpha,\n} = | 0_{\n} \rangle \langle \alpha_\n |,
\;\;\; \hat{a}^{\dagger}_{\alpha,\n} = | \alpha_{\n} \rangle \langle 0_{\n} |\quad (\alpha \neq 0 ),
\ee
with the closure relation
\be
\sum_{\alpha=0}^{M} |\alpha_{\n}\rangle\langle\alpha_{\n}
| = I
\ee
and the orthonormality condition
\be 
\langle \alpha_{\n} | \beta_{\n} \rangle = \delta_{\alpha\beta},
\;\;\; \alpha, \beta = 0, 1, \ldots, M.
\ee
These operators obey the hard-core bosonic commutation relations
\begin{subequations}
\label{Eq:HCOperator_Comm}
\bea
\label{Eq:HCproperty}
\hat a_{\alpha,{\n}} \hat a_{\beta,{\n}} &=& 
\hat a_{\alpha,{\n}}^{\dag}
\hat a_{\beta,{\n}}^\dag = 0,\\
\label{Eq:anticom}
\hat a_{\alpha,{\n}} \hat a_{\alpha,{\n}}^\dag &=& 1 -
\sum_{\gamma =1}^M \hat a_{\gamma,{\n}}^\dag 
\hat a_{\gamma,{\n}},\\
\label{Eq:com}
{} [ \hat a_{\alpha, {\n}},\hat a_{\beta,{\m}}^\dag ]&=&
[ \hat a_{\alpha,{\n}},\hat a_{\beta,{\m}}]=0
\quad ({\n} \neq {\m}),
\eea
\end{subequations}
where Eqs. (\ref{Eq:HCproperty}) and (\ref{Eq:anticom}) 
prohibit double occupancy at a single 
site even for different species---hence the nomenclature ``hard core.''
Equation (\ref{Eq:com}) states that any two operators defined at different
sites commute---hence the name ``boson.''
As usual, $\hat N_{\alpha,{\n}} = 
\hat a_{\alpha,{\n}}^\dag \hat a_{\alpha,{\n}}$ is 
the number operator whose eigenvalues are 0 (vacant) and 1 (occupied).

To each state of the system, we can associate the state vector
$|P;t\rangle$, considered an element of Fock space,
\be
|P;t\rangle = \sum_{ \{ N \} } P(\{ N \};t) |\{ N \}\rangle,
\ee
where $|\{ N \}\rangle$ is the microstate defined in Eq. (\ref{Eq:microstate}).
The microstate can be written in terms of hard-core boson operators as
\be
\label{Eq:state_vector_of_HC}
\ba
| \{ N \} \rangle &= \prod_{\n} \prod_{\alpha=1}^M ( \had{\alpha}{\n}
)^{N_{\alpha,\n}} | \{ 0 \} \rangle,\\
\langle \{ N \} | &= \langle \{ 0 \}| \prod_{\n} \prod_{\alpha=1}^M
(\ha{\alpha}{\n} )^{N_{\alpha,\n} },
\ea
\ee
where $N_{\alpha,\n}$ is the eigenvalue of the number operator
corresponding to the eigenstate $\{ N \} \rangle$
($\hat N_{\alpha,\n} | \{ N \} \rangle = N_{\alpha,\n} | \{ N \} \rangle$).
Using hard-core boson operators, the projection state  can be written as
\be
\langle \cdot | =\langle \{ 0 \}| \prod_{\n} \left ( 
1 + \sum_{\alpha=1}^M \ha{\alpha}{\n} \right ).
\label{Eq:HC_projection}
\ee

Now we introduce
the generating function $F( \{ \bar \varphi \} ;t )$, defined as
\be
\ba
F ( \{ \bar \varphi \} ;t ) &= \langle \{ \bar \varphi \} | P
;t \rangle,
\label{Eq:Gen_ftn}
\ea
\ee
where 
\be
\ba
\langle \{ \bar \varphi \} | &\equiv \bigotimes_{\n} \left ( \langle 0_{\n} | +
\sum_{\alpha = 1}^M
\langle \alpha_{\n} |f(\bar \varphi_{\alpha,{\n}})  \right )
\\
& =
\langle \cdot | \prod_{\n} \left (1 
+ \sum_{\alpha=1}^M [ f(\bar \varphi_{\alpha,{\n}} ) - 1]
\hat a_{\alpha,{\n}}^\dag \hat a_{\alpha,{\n}} \right ),
\ea
\label{Eq:Gen_ftn_bra}
\ee
and $f$ is an analytic function of its argument.
$F$ is the generating function in that 
by differentiating $F$ with respect to $f$ and setting $f = 1$, we can 
find all kinds of correlation functions of the particle density.
For example,
\be
\langle N_{\alpha,\n} \rangle = \left . 
\frac{\partial}{\partial f(\bar \varphi_{\alpha,\n} )}
F(\{ \bar \varphi \} ;t) 
\right |_{f=1} = \langle \cdot | \had{\alpha}{\n} \ha{\alpha}{\n} 
| P ;t \rangle.
\ee
The exact form of $f$ does not matter. In what follows,
we mainly use three kinds of functions,
\begin{subequations}
\label{Eq:f}
\bea
f(\bar \varphi) &=& \bar \varphi, \label{Eq:f_symmetric}\\
f(\bar \varphi) &=& 1 + \bar \varphi,\label{Eq:f_boson_related}\\
f(\bar \varphi) &=& \exp(\bar \varphi),\label{Eq:f_prev_form_related}
\eea
\end{subequations}
and these will be used in different contexts.

By differentiating Eq. (\ref{Eq:Gen_ftn}) with respect to $t$, 
we obtain
\be
\frac{\partial}{\partial t} F(\{ \bar \varphi\};t) = - \langle \{ \bar \varphi \} |
\hat H | P ;t \rangle.
\label{Eq:time_evolve_gen}
\ee
Since our main goal is to find a path integral
representation of $F(\{\bar \varphi\};t)$ from Eq.
(\ref{Eq:time_evolve_gen}), we need to find a linear partial
differential equation about $F(\{ \bar \varphi \};t)$.
This can be achieved using the following properties of $\langle \{ \bar \varphi \} |$:
\begin{subequations}
\label{Eq:op_to_diff}
\bea
\label{Eq:op_to_diff_1}
\langle \{ \bar \varphi \} | \hat a_{\alpha,{\n}}^\dag  &=& f(\bar
\varphi_{\alpha,{\n}}) \hat V_\n
\langle \{ \bar \varphi \} |,\\
\label{Eq:op_to_diff_2}
\langle \{ \bar \varphi \} | \hat  a_{\alpha,{\n}} &=&
\frac{\partial}{\partial  f(\bar \varphi_{\alpha,{\n}})} \langle
\{ \bar \varphi \} |,\\
\label{Eq:op_to_diff_3}
\langle \{ \bar \varphi \} | \hat a_{\alpha,{\n}}^\dag \hat
a_{\beta,{\n}} &=&  f(\bar \varphi_{\alpha,{\n}})
\frac{\partial}{\partial  f(\bar \varphi_{\beta,{\n}})}
\langle \{ \bar \varphi \} |,\\
\label{Eq:op_to_diff_4}
\langle \{ \bar \varphi \} | \hat a_{\alpha,{\n}} \hat a_{\beta,{\n}}^\dag &=& 
\delta_{\alpha\beta} \hat V_\n \langle \{ \bar \varphi \} |,
\eea
\end{subequations}
where \[\hat V_\n  \equiv \left ( 1 - \sum_{\beta=1}^M
\frac{\partial}{\partial \ln f(\bar \varphi_{\beta,{\n}}) }\right )\]
is the projection operator to the vacuum state at site $\n$.
Relations (\ref{Eq:op_to_diff}) yield the partial differential equation 
for the generating function,
\be
\frac{\partial}{\partial t} F(\{\bar \varphi \};t) = -  {\cal L}\left
(\{ \bar \varphi \},
\left \{\frac{\partial}{\partial \bar \varphi }\right \} \right )
F(\{\bar \varphi \};t),
\label{Eq:main_equation}
\ee
where  $\cal L$ takes the normal-ordered form, that is, 
all $\bar \varphi$'s are located to the left side of 
any $\partial/\partial \bar \varphi $.
We call $\cL$ an evolution operator.
Since this is a linear equation, we can write the path integral
solution of $F(\{\bar \varphi \};t)$.
The path integral solution of Eq.  (\ref{Eq:main_equation}) with any of
the prescription Eqs. (\ref{Eq:f}) is
\be
F (\{\bar \xi \};t) = \int\prod_{{\alpha, \n}} 
{ \frac{d \eta_{{\alpha,\n}} d \bar \eta_{{\alpha,\n}}}{2 \pi
    i}e^{-\eta_{\alpha,\n}\bar \eta_{\alpha,\n}}} 
F(\{\bar \eta\};0) T^{\{\bar \xi\}}_{\{\eta\}}(t),
\ee with
\be
\label{Eq:Tetaxi}
T^{ \{\bar \xi\} }_{ \{\eta\} }(t) = \int {\cal D} \{ \bar \varphi \} {\cal D}
\{ \varphi\} e^{ - S (\{\bar \varphi\}, \{\varphi\},t) + \{ \bar \xi
  \} \cdot  \{\varphi(t)\} + \{\eta\} \cdot \{\bar \varphi(0) \}},
\ee
where
\begin{equation}
\label{Eq:HC_genftn_action}
S = \int_0^t  dt \left [\sum_{\alpha,{\n}}\bar 
\varphi_{\alpha,{\n}} \frac{\partial}{\partial t}
\varphi_{\alpha,{\n}}
+ {\cal L}(\{\bar \varphi(t)\}, \{\varphi(t)\} )\right ]
\end{equation}
and
\begin{equation}
\{ \bar \xi \} \cdot \{ \eta \} = \sum_{\alpha,\n} \bar
\xi_{\alpha,\n} \eta_{\alpha,\n}.
\end{equation}
Taking a continuum limit and keeping the most relevant terms, we
arrive at mesoscopic action which is equivalent to the microscopic
master equation. The long-time properties are extracted by studying
the action using renormalization-group theory.

\section{Equivalence between stochastic systems}
The equivalence between stochastic systems  has been usually studied 
using similarity transformation for single-species 
reaction-diffusion systems \cite{HOS95}. 
This section shows the equivalence between stochastic systems
using the evolution equation of the generating function 
in Eq. (\ref{Eq:main_equation}) with the prescription 
Eq. (\ref{Eq:f_boson_related}) instead of using the similarity transformation.
The key mechanism to deduce the equivalence between stochastic systems 
is the rescaling of the field
$\bar \varphi$ (see below), which is simple enough to be applicable 
to higher-dimensional systems and multispecies systems.

To begin with, we consider a single-species reaction-diffusion 
model which is defined on a $d$-dimensional hypercubic lattice
with  diffusion, pair annihilation, coalescence, death,
and single-particle branching.  Particles move with a diffusion
constant $D$. When two particles form a nearest-neighbor pair, both
of them are  annihilated with rate $\lambda$ or one of them is removed with
rate $\eta$. Additionally, a single particle is annihilated spontaneously 
with rate $\delta$ and a particle-vacant pair becomes a particle-particle
pair with rate $\sigma$.
The dynamics are summarized in Table \ref{Table:DPrates}.
When $\delta=\sigma=0$, this model corresponds to the single-species
annihilation and coalescence model. 
The model with $D=\lambda=\eta=0$ is the well-known contact process
\cite{H74}.
When all processes are
present, this model is known to show an absorbing phase transition
which shows the same critical behavior as that of 
the directed percolation \cite{H00}.

The \itH of this reaction-diffusion system is
$\hat H = \sum_\n ( \hat H_\n^D + \hat H_\n^\lambda +
\hat H_\n^\eta + \hat H_\n^\delta + \hat H_\n^\sigma )$ 
with 
\begin{subequations}
\label{Eq:DP_Ham1}
\bea
&&\hat H_\n^D = D \sum_{i=1}^d\big [  \rn\n( 1 - 
\rn{\n + \e_i}) - 
\hat a_\n \hat a_{\n+\e_i}^\dag 
+( 1 - \rn\n ) \rn{\n + \e_i}
- \hat a_\n^\dag \hat a_{\n + \e_i} \big ],
\label{Eq:DP_Ham_diff}
\\
&&\hat H_\n^\lambda = \lambda \sum_{i=1}^d ( 
\rn\n \rn{\n + \e_i}
- \hat a_\n  \hat a_{\n + \e_i} ),
\label{Eq:DP_Ham_ann}
\\
&&\hat H_\n^\eta = \frac{\eta}{2} \sum_{i=1}^d (2 
\rn\n \rn{\n + \e_i} - 
\rn\n  \hat a_{\n + \e_i}
- \hat a_\n  \rn{\n + \e_i}),
\label{Eq:DP_Ham_coal}
\\
&&\hat H_\n^\delta = \delta ( \rn\n  - \hat a_\n ),
\label{Eq:DP_Ham_death}
\\
&&\hat H_\n^\sigma = \frac{\sigma}{2} \sum_{i=1}^d \big [ \rn\n
( 1 - \rn{\n + \e_i} ) - 
\rn\n \hat a_{\n + \e_i}^\dag+ ( 1 - \rn\n ) 
\rn{\n + \e_i}  - \hat a_\n^\dag
\rn{\n + \e_i} \big ],
\label{Eq:DP_Ham_branch}
\eea
\end{subequations}
where $\rn\n \equiv \hat a_\n^\dag \hat a_\n$ is the number operator and 
we drop the unnecessary species index.
The operators satisfy the commutation relations Eqs.
(\ref{Eq:HCOperator_Comm}) with $M=1$.
Following the procedure explained in Sec. \ref{Sec:formalism},
we can find the evolution operator $\cL$ with the prescription Eq.
(\ref{Eq:f_boson_related}).
If we write $\cL$ as
$\cL =
\sum_{\n}\sum_{i=1}^d \cL_{\n,i}$, we obtain 
\be
\ba
\cL_{\n,i} &= ( 2  D + \delta )
\bar \varphi_\n\frac{\partial}{\partial \bar \varphi_{\n}} - 
\left ( D + \frac{\sigma}{2} \right ) 
\left ( \bar \varphi_{\n} \frac{\partial}{\partial
\bar \varphi_{\n+\e_i}} + \bar \varphi_{\n+\e_i} \frac{\partial}{\partial
\bar \varphi_{\n}} \right ) \\
&+ \left ( \lambda + \frac{\eta}{2} + \frac{\sigma}{2} \right ) 
( \bar \varphi_{\n} + \bar \varphi_{\n+\e_i} ) \frac{\partial^2}{\partial
\bar \varphi_{\n} \partial \bar \varphi_{\n+\e_i}}\\
&+ ( \lambda + \eta + \sigma - 2 D  ) \bar \varphi_{\n} \bar \varphi_{\n+\e_i} 
\frac{\partial^2}{\partial
\bar \varphi_{\n} \partial \bar \varphi_{\n+\e_i}}
+ \left (  D + \frac{\sigma}{2} \right ) 
\left ( \bar \varphi_\n^2 + \bar \varphi_{\n + \e_i}^2 \right )
\frac{\partial^2}{\partial
\bar \varphi_{\n} \partial \bar \varphi_{\n+\e_i}}
\label{Eq:DP_EVO}\\
&+ \frac{\sigma}{2}  \bar \varphi_\n
\bar \varphi_{\n+\e_i}\left [ (\bar \varphi_\n + \bar \varphi_{\n+\e_i} )
\frac{\partial^2}{\partial \bar \varphi_{\n} \partial \bar \varphi_{\n+\e_i}}
- 
\left ( \frac{\partial}{\partial \bar \varphi_{\n}} + \frac{\partial}{\partial
\bar \varphi_{\n+\e_i}} \right ) \right ]. 
\ea\ee
If we use an uncorrelated initial condition with density $\rho_0$,
the initial-state vector and the initial generating function can be written as
\be
\ba
| P ;0 \rangle &= \prod_\n \left [ ( 1 - \rho_0 ) + \rho_0 \hat a_\n^\dag \right ] |
0 \rangle,\\
F(\{\bar \varphi \};0 ) &= \prod_\n ( 1 + \rho_0 \bar \varphi_\n ).
\ea
\label{Eq:DP_initial}
\ee
Hence Eqs. (\ref{Eq:DP_EVO}) and (\ref{Eq:DP_initial}) along with 
Eq. (\ref{Eq:main_equation}) fully specify the above
reaction-diffusion system. 

Let us assume that the solution of
Eq. (\ref{Eq:main_equation}) is written as 
$F(\{ \bar \varphi \}, D,\lambda,\eta,\delta,\sigma,\rho_0;t)$.
Rescaling the field $\bar \varphi = \mu \bar \varphi'$ ($\mu > 0$), 
$F$ is modified to
\be
F(\{\bar \varphi \},  D,\lambda,\eta,\delta,\sigma,\rho_0;t) = 
F(\{\bar \varphi' \}, \tilde D, \tilde \lambda,\tilde \eta,\tilde \delta,\tilde \sigma,\mu \rho_0;t),
\label{Eq:Equiv}
\ee
where the relations between parameters with and without a tilde  
are found by setting
$\bar \varphi = \mu \bar \varphi'$ in Eq. (\ref{Eq:DP_EVO}),
which read
\be
\ba
\tilde \sigma &= \mu \sigma,\\
2  \tilde D + \tilde \delta &= 2  D  + \delta,\\
\tilde D + \frac{\tilde \sigma}{2} &= D + \frac{\sigma}{2},\\
\tilde \lambda + \frac{\tilde \eta}{2} + \frac{\tilde \sigma}{2} &=
\frac{1}{\mu}\left (\lambda + \frac{\eta}{2} + \frac{\sigma}{2}\right ),\\
\tilde \lambda + \tilde \eta + \tilde \sigma - 2 \tilde D  &= \lambda + \eta + \sigma - 2 D .
\ea 
\label{Eq:prime_unprime}
\ee
Since $F$ on the right-hand side of Eq. (\ref{Eq:Equiv}) can be regarded as 
a solution of Eq. (\ref{Eq:main_equation}) with tilded rates and initial 
density $\mu \rho_0$, the two systems connected by 
Eq. (\ref{Eq:prime_unprime}) share the same generating function.
The relation of the correlation function can be found by differentiating
$F$ with $\bar \varphi$, which reads
\be
C_k(\{{\bm x}\},D,\lambda,\eta,\delta,\sigma,\rho_0;t)
= \mu^{-k} \tilde C_k(\{{\bm x}\},\tilde D,\tilde \lambda,\tilde \eta,
\tilde \delta, \tilde \sigma,\mu \rho_0;t),
\ee
where $C_k$ ($\tilde C_k$) is the $k$-point correlation functions of
the reaction-diffusion systems with the untilded (tilded) transition rates.
Thus, for arbitrary $\mu$, which ensures all tilded parameters are non-negative,
we can find the equivalent stochastic systems to the system with untilded
parameters.

Let us find the equivalent systems to the single-species 
pair annihilation model ($\eta=\sigma=\delta=0$).
For given $\mu$, the tilded rates are found as
\be
\tilde D = D, \tilde \sigma = 0, \tilde \delta = 0,
\tilde \eta = 2 \left ( 1 - \frac{1}{\mu} \right ) \lambda,
\tilde \lambda = \left ( \frac{2}{\mu}  - 1 \right ) \lambda,
\label{Eq:sim_ann}
\ee
where $1\le \mu \le 2$ should be satisfied to have the physical meaning.
If we choose $\mu = 2$ in Eq. (\ref{Eq:sim_ann}), the tilded rates 
become $\tilde D=D$, $\tilde \eta = \lambda$, and 
$\tilde \delta=\tilde \sigma=\tilde \lambda =0$,
which are the transition rates of the single-species coalescence model with
the initial density $\tilde \rho_0 = 2 \rho_0$ (of course $\rho_0$ should
not be larger than $\frac{1}{2}$). 
Hence 
all kinds of correlation functions of the pair annihilation 
and coalescence models are related to one another in any dimension.
This method also reproduces all results in Ref. \cite{HOS95} 
regarding the equivalence of stochastic systems by adjusting $\mu$ and also finds
the initial condition relation.

Next, we apply the generating function method to 
find the equivalence between three-particle
annihilation models. 
To our knowledge, this equivalence has not been studied
in the literature, although the full 
renormalization-group study of this model can be found in Ref. 
\cite{Lee94}.
The dynamics of the three-particle annihilation model is summarized in Table
\ref{Table:3A.goto}.
The procedure to find the equivalence relations is the same as those explained
above. First, we find the {\itH} and the corresponding evolution operator.
The corresponding {\itH} is $\hat H = \sum_\n \sum_{i=1}^d\left 
( \hat H_{\n,i}^D + \hat H_{\n,i}^{\lambda_1}+ \hat H_{\n,i}^{\lambda_2}
+ \hat H_{\n,i}^{\lambda_3}\right )$
with 
\begin{subequations}
\label{Eq:DP_Ham2}
\bea
&&\hat H_{\n,i}^D = D \big [  \rn{\n}( 1 - 
\rn{\n + \e_i}) - 
\hat a_\n \hat a_{\n+\e_i}^\dag  
+( 1 - \rn\n ) \rn{\n + \e_i}
- \hat a_\n^\dag \hat a_{\n + \e_i} \big ],
\\
&&\hat H_{\n,i}^{\lambda_1} = \frac{\lambda_1}{3} \big [ 
3 \rn\n \rn{\n + \e_i}
\rn{\n + 2 \e_i}
- \hat a_\n  \rn{\n + \e_i}\rn{\n + 2 \e_i} 
- \rn\n  \rn{\n + \e_i} \hat a_{\n + 2 \e_i} 
- \rn\n  \hat a_{\n + \e_i} \rn{\n + 2 \e_i} 
\big ],
\\
&&\hat H_{\n,i}^{\lambda_2} = \frac{\lambda_2}{3} \big [ 
3 \rn\n \rn{\n + \e_i}
\rn{\n + 2 \e_i}
- \hat a_\n  \hat a_{\n + \e_i}\rn{\n + 2 \e_i} 
- \hat a_\n  \rn{\n + \e_i} \hat a_{\n + 2 \e_i} 
- \rn\n  \hat a_{\n + \e_i} \hat a_{\n + 2 \e_i} 
\big ],
\\
&&\hat H_{\n,i}^{\lambda_3} = {\lambda_3} \big [
 \rn\n \rn{\n + \e_i}
\rn{\n + 2 \e_i}
- \hat a_\n  \hat a_{\n + \e_i}\hat a_{\n + 2 \e_i} 
\big ],
\eea
\end{subequations}
and the evolution operator is $\cL = \sum_\n \sum_i \cL_{\n,i}$ with
\be
\ba
\cL_{\n,i} &=  
D(\bvp\n - \bvp{\n+\e_i})\left (\vp\n - \vp{\n+\e_i} \right ) + D 
(\bvp\n - \bvp{\n+\e_i})^2 \vp\n \vp{\n+\e_i}\\
&+\frac{1}{3}(\lambda_1 + 2 \lambda_2 + 3 \lambda_3 )
(\bvp\n+\bvp{\n+\e_i} + \bvp{\n+2 \e_i} ) \vp\n \vp{\n+\e_i}\vp{\n+2 \e_i}\\
&+\frac{1}{3} ( 2 \lambda_1 + 3 \lambda_2 + 3 \lambda_3) 
( \bvp\n \bvp{\n+\e_i} + \bvp{\n+\e_i} \bvp{\n+2 \e_i} +
\bvp{\n+2 \e_i} \bvp\n )  \vp\n \vp{\n+\e_i}\vp{\n+2 \e_i}\\
&+(\lambda_1 + \lambda_2 + \lambda_3) \bvp\n \bvp{\n+\e_i} \bvp{\n+2 \e_i}
\vp\n \vp{\n+\e_i}\vp{\n+2 \e_i} .
\ea
\ee
By rescaling $\bvp{} = \mu\bvp{}'$, we find the parameter relations
\be
\ba
&\tilde D = D, \tilde \lambda_1 + 2 \tilde \lambda_2 + 3 \tilde \lambda_3
= \mu^{-2} (\lambda_1 +2 \lambda_2 + 3 \lambda_3),\\
&2 \tilde \lambda_1 + 3 \tilde \lambda_2 + 3 \tilde \lambda_3
= \mu^{-1} (2 \lambda_1 +3 \lambda_2 + 3 \lambda_3),\\
&\tilde \lambda_1 +  \tilde \lambda_2 +  \tilde \lambda_3
= \lambda_1 + \lambda_2 +  \lambda_3 .
\ea
\ee
First consider $3 A \rightarrow 0$ only ($\lambda_1 = \lambda_2  = 0$). 
Unfortunately, the only
possible value of $\mu$ is 1, which means there is no equivalent 
procedure to $3A \rightarrow 0$.
However, if we set $\lambda_2 = \lambda_3 = 0$, we find
\be
\ba
\tilde \lambda_1 =\lambda_1 \left ( 3 - \frac{2}{\mu}\right ),
\tilde \lambda_2 = \lambda_1 \frac{-1 + 4 \mu - 3 \mu^2}{\mu^2},
\tilde \lambda_3 = \lambda_1 (1 - \mu^{-1})^2,
\ea
\ee
where the valid range of $\mu$ is $2/3 \le \mu \le 1$.

Up to now, we have found the equivalence for single-species 
reaction-diffusion systems, but the formalism explained above can be extended 
to multispecies problems. As an example, let us consider
a $d$-dimensional two-species annihilation model. 
There are two kinds of species, $A$ and $B$. Both particles have
the same  diffusion constant $D$.
When two different species form a pair, one of the following reactions 
may occur.
Both particles are removed, or only one particle
is annihilated. For the-one particle annihilation process, there are 
two possibilities. The remaining species stays where it was or moves
to the other site. 
For these dynamics, we assign transition rates $\lambda$ (pair annihilation),
$\eta$ (one-particle annihilation without location change), and $\zeta$ 
(one-particle annihilation with location change).
The dynamics is summarized in Table \ref{Table:two-species}.

The {\itH} for this model is $\hat H = \sum_{\n} \sum_{i=1}^d \hat H_{\n,i}$ with 
\begin{subequations}
\label{Eq:two-species}
\begin{eqnarray}
&&\hat H^D_{\n,i} = D ( \A\n \vac{\n+\e_i} - \hat a_\n \hat a_{\n+\e_i}^\dag)
 + D ( \B\n \vac{\n+\e_i} - \hat b_\n \hat b_{\n+\e_i}^\dag)\\
&&\hspace{1cm}+ D ( \A{\n+\e_i} \vac\n - \hat a_{\n+\e_i} \hat a_\n^\dag)
 + D ( \B{\n+\e_i} \vac\n - \hat b_{\n+\e_i} \hat b_\n^\dag),\nonumber\\
&&\hat H^\lambda_{\n,i} = \lambda (\A\n \B{\n+\e_i} -\hat  a_\n \hat b_{\n + \e_i} 
+\B\n \A {\n+\e_i} - \hat b_\n \hat a_{\n + \e_i} ),\\
&&\hat H^\eta_{\n,i} = \eta (\A\n \B{\n+\e_i} -\hat a_\n \B{\n + \e_i} 
+\A\n \B{\n+\e_i} -\A\n \hat b_{\n + \e_i} \\
&&\hspace{1cm} +\B\n \A {\n+\e_i} - \hat b_\n \A{\n + \e_i}  
+\B\n \A {\n+\e_i} - \B\n \hat a_{\n + \e_i} ),\nonumber\\
&&\hat H^\zeta_{\n,i} = \zeta (\A\n \B{\n+\e_i} -\hat b_\n^\dag \hat a_\n \hat b_{\n + \e_i} 
+\A\n \B{\n+\e_i} -\hat a_\n \hat a^dag_{\n + \e_i} \hat b_{\n + \e_i} \\
&&\hspace{1cm} +\B\n \A {\n+\e_i} - \hat a_\n^\dag \hat b_\n \hat a_{\n + \e_i} 
+\B\n \A {\n+\e_i} - \hat b_\n \hat b_{\n + \e_i}^\dag \hat a_{\n + \e_i} ),
\nonumber
\end{eqnarray}
\end{subequations}
where $\A\n~(\B\n)$ is the number operator for species $A$ ($B$),
and $\vac\n$ is the vacuum projection operator with the form
$\vac\n = 1 - \A\n - \B\n$. We use a different symbol for the different
species and drop the species indices. 
The commutation relations for the operators 
in Eq. (\ref{Eq:two-species}) correspond to the $M=2$ case 
of Eq. (\ref{Eq:HCOperator_Comm}). 
The evolution operator takes the form $\cL = \sum_{\n}\sum_{i=1}^d \cL_{\n,i}$ with
\be
\ba
\cL_{\n,i} &= D 
\left [\left ( \bvp\n - \bvp{\n + \e_i} \right ) \left ( \vp\n - \vp{\n + \e_i} \right ) + 
\left (\bvps\n - \bvps{\n + \e_i} \right ) \left ( \vps\n - \vps{\n + \e_i}
\right )\right ] \\
&+ D\left [ ( \bvp\n - \bvp{\n+\e_i} )^2 \vp\n \vp{\n+\e_i} + 
( \bvps\n - \bvps{\n+\e_i} )^2 \vps\n \vps{\n+\e_i} \right ]\\
&+ D \left ( \bvp\n \bvps{\n+\e_i} + \bvps\n \bvp{\n+\e_i} \right )
\left ( \vp\n \vps{\n+\e_i} + \vps\n \vp{\n+\e_i} \right )\\
&+\left ( D - \eta \right ) 
\left [ \left ( \bvp\n + \bvps{\n + \e_i} \right ) \vps\n \vp{\n + \e_i}
+ \left ( \bvps\n + \bvp{\n + \e_i} \right ) \vp\n \vps{\n + \e_i}
\right ]\\
&+( \zeta + 2\eta +\lambda - d D ) 
\left [ \left ( \bvp\n + \bvps{\n + \e_i} \right ) \vp\n \vps{\n + \e_i}
+ \left ( \bvps\n + \bvp{\n + \e_i} \right ) \vps\n \vp{\n + \e_i}
\right ]\\
&+( 2 \zeta + 2 \eta + \lambda - 2 d D ) 
\left ( \bvp\n \bvps{\n + \e_i} \vp\n \vps{\n + \e_i}
+ \bvps\n \bvp{\n + \e_i} \vps\n \vp{\n + \e_i} \right ),
\ea
\ee
where $\bar \varphi (\bar \psi)$ corresponds to species $A$ ($B$).
By rescaling, we find the relations
\be
\ba
&\tilde D = D,\;\;
\tilde D - \tilde \eta = \frac{1}{\mu} ( D - \eta ),\\
&\tilde \zeta + 2 \tilde \eta + \tilde \lambda -  \tilde D
= \frac{1}{\mu}(\zeta + 2\eta +\lambda -  D),\\
&2 \tilde \zeta + 2 \tilde \eta + \tilde \lambda  - 2  \tilde D
= 2 \zeta + 2 \eta + \lambda - 2  D,
\ea
\ee
or
\be
\ba
\tilde D &=D,\;\;
\tilde \zeta = \zeta + \frac{\mu-1}{\mu} (\zeta + 2 \eta + \lambda -  D),\\
\tilde \lambda &= \lambda  + \frac{2 - 2 \mu}{\mu} (  \zeta +  \eta + \lambda),
\;\;
\tilde \eta = \frac{\eta}{\mu} + \frac{\mu-1}{\mu} D.
\ea\ee
If we set $\eta = \zeta = 0$, we can find 
\be
\tilde D = D, \tilde \eta = \frac{\mu-1}{\mu} D,\;\; \tilde \lambda = \frac{2-\mu}{\mu} \lambda,
\;\;\tilde \zeta = \frac{\mu - 1}{\mu} ( \lambda - D ).
\ee
If $\lambda \ge D$,  we obtain equivalent stochastic
processes with the pair annihilation model $A+B\rightarrow \emptyset$
for $1 \le \mu \le 2$.
Other equivalent relations can be found by adjusting $\mu$ for 
different values of  untilded parameters.

\section{Summary and Discussion}
In summary, we have presented the path integral representation of the
generating function for classical stochastic exclusive systems.
Using this formalism, we have shown the equivalences of some stochastic
systems of single- and two-species models.
Furthermore, since our formalism is insensitive to the
choice of the function $f$ in Eq. (\ref{Eq:Gen_ftn_bra}), we can have
different action forms for one system depending on the choice of
$f$. Although they may have different forms, the long-time behavior of
physical quantities derived from each action form should be the
same. For example, if we use the prescription Eq. (\ref{Eq:f_prev_form_related})
in Eq. (\ref{Eq:Gen_ftn_bra}), the action is
equivalent to the action with the choice of $f(\bar \varphi) = \bar \varphi$ in
Eq. (\ref{Eq:Gen_ftn_bra}) and with the change of fields, such as $\bar
\varphi \mapsto e^{\bar \varphi}$ and $\varphi \mapsto e^{- \bar
\varphi} \varphi$, performed.
In fact, if  we use the function $f(\bar \varphi) =
e^{\bar \varphi}$ and expand the
function $f$ in terms of $\bar \varphi$ keeping only terms 
up to $\bar \varphi^2$ order, we obtain
the same action as the one derived in Ref. \cite{Park2000}.
In many cases, any term which contains $\bar \varphi$ higher than
second order is irrelevant. The method introduced in Ref.
\cite{Park2000} is legitimized by the force of the renormalization-group argument.
In some cases, usually in reaction-diffusion systems, the
canonical dimension of $\bar \varphi$ is 0 \cite{Lee94,Lee95}, so it is not clear whether
we can neglect the higher-order terms after expanding $f$.
In this situation, one should use Eq. (\ref{Eq:f_symmetric}) or Eq.
(\ref{Eq:f_boson_related}) to ensure the completeness of the
formalism. 
As an example, let us consider again the model described by the
reaction given in Table \ref{Table:DPrates} with $\eta = \delta =
\sigma = 0$. If we use Eq. (\ref{Eq:f_boson_related}), the action
becomes (up to relevant terms)
\be
S = \int dt  d^d x \left \{ \bar \varphi 
\left ( \partial_t - D \nabla^2 \right ) \varphi + 2 \lambda \bar \varphi
\varphi^2 + \lambda \bar \varphi^2 \varphi^2 \right \},
\label{Eq:AA_hard_core}
\ee
where we neglect the term $D \left ( \nabla \bar \varphi \right )^2
\varphi^2$ due to its irrelevance.
Equation (\ref{Eq:AA_hard_core}) is exactly the same as the action 
in Ref. \cite{Lee94} obtained using a bosonic formalism.
Since we expect that boson and hard-core boson 
annihilation models should belong to the same universality class,
this coincidence is not surprising.
If we perform the change of fields explained above, 
and keeping terms up to $O(\bar \varphi^2)$,
the action takes the form
\be
\ba
S' =& \int dt d^d x \{ \bar \varphi ( \partial_t - D \nabla^2 ) \bar
\varphi - D \varphi ( \nabla \bar \varphi)^2 \\
&\hspace{1cm}+ 2 \lambda \bar \varphi \varphi^2 - 2 \lambda \bar \varphi^2
\varphi^2  \},
\label{Eq:AA_HC_realnoise}
\ea
\ee
which can also be found by the method introduced in Ref. \cite{Park2000}.
These two actions, Eq. (\ref{Eq:AA_hard_core}) and
Eq. (\ref{Eq:AA_HC_realnoise}), look different, although they
describe the same system. This is due to the dropping of
higher-order terms of $\bar \varphi$. To describe the system properly,
we have to keep all the higher-order terms of  $\bar \varphi$ which
are marginal. Then, two actions will give the same long-time behavior
of physical quantities.

Although we have concentrated on reaction-diffusion systems,
the applicability of this formalism is not restricted to 
reaction-diffusion systems. 
We found Langevin equations for DLG \cite{Zia} and other lattice-gas-type 
models. This will be published elsewhere \cite{unpub}.

\section*{Acknowledgments}
This research was supported by the Catholic University of Korea
research fund 2004 and Grant No. R05-2003-000-12072-0 from the BRP
program of the KOSEF.

\begin{table}[h]
\begin{center}
\caption{\label{Table:DPrates} Reaction-diffusion processes of single
species and their rates.}
\begin{ruledtabular}
\begin{tabular}{ccl}
\text{Diffusion}& $A\emptyset \leftrightarrow \emptyset A$ & \text{with
rate }$D$\\
\text{Pair annihilation}&$ A A \rightarrow \emptyset \emptyset$  &
\text{with rate }$\lambda$\\
\text{Coalescence} & $ A A \rightarrow A \emptyset$  &
\text{with rate }$ \eta/2$ \\
\text{Coalescence} & $ A A \rightarrow \emptyset A$
&\text{with rate } $ \eta/2 $\\
\text{Death} &  $A   \rightarrow \emptyset$
&\text{with rate } $\delta$ \\
\text{Branching}  &  $\emptyset A \rightarrow A A $
&\text{with rate } $\sigma/2$ \\
\text{Branching}  & $ A \emptyset  \rightarrow A A $&
\text{with rate } $\sigma/2 $\\
\end{tabular}
\end{ruledtabular}
\end{center}
\end{table}
\begin{table}[h]
\begin{center}
\caption{\label{Table:3A.goto} Reaction-diffusion processes of 
the three-particle annihilation model and their rates.}
\begin{ruledtabular}
\begin{tabular}{lc}
$A\emptyset \leftrightarrow \emptyset A$ & \text{with
rate }$D$\\
$AAA\rightarrow A\emptyset A, \emptyset A A, \text{ or } A A\emptyset$
&\text{with rate} $\lambda_1/3$\\
$AAA\rightarrow A\emptyset \emptyset, \emptyset A \emptyset, \text{ or }  \emptyset\emptyset A$
&\text{with rate} $\lambda_2/3$\\
$AAA\rightarrow \emptyset \emptyset \emptyset $&
\text{with rate} $\lambda_3$\\
\end{tabular}
\end{ruledtabular}
\end{center}
\end{table}
\begin{table}[ht]
\caption{\label{Table:two-species} Dynamics of the two-species 
annihilation model and respective rates.}
\begin{ruledtabular}
\begin{tabular}{lc}
$A\emptyset\leftrightarrow \emptyset A$&with rate $D$\\
$B\emptyset\leftrightarrow \emptyset B$&with rate $D$\\
$AB \rightarrow \emptyset\emptyset$&with rate $\lambda$\\
$BA \rightarrow \emptyset\emptyset$&with rate $\lambda$\\
$AB \rightarrow A\emptyset \text{ or } \emptyset B$&with rate $\eta$\\
$BA \rightarrow B\emptyset \text{ or } \emptyset A$&with rate $\eta$\\
$AB \rightarrow B\emptyset \text{ or } \emptyset A$&with rate $\zeta$\\
$BA \rightarrow A\emptyset \text{ or } \emptyset B$&with rate $\zeta$\\
\end{tabular}
\end{ruledtabular}
\end{table}
\end{document}